\documentclass[sigconf]{acmart}

\AtBeginDocument{%
  }

\setcopyright{acmlicensed}
\copyrightyear{2025}
\acmYear{2025}
\acmDOI{XXXXXXX.XXXXXXX}
\acmConference[UMAP]{33rd ACM International Conference on User Modeling, Adaptation and Personalization}{June 16--19,
  2025}{NYC, NY}
\acmISBN{978-1-4503-XXXX-X/2018/06}
\usepackage[font=small]{}   

\begin{document}

\title{Towards Intelligent VR Training: A Physiological Adaptation Framework for Cognitive Load and Stress Detection}

\author{Mahsa Nasri}
\email{nasri.m@northeastern.edu}
\orcid{0000-0003-3467-5279}

\affiliation{%
  \institution{Northeastern University}
  \city{Boston}
  \state{Massachusets}
  \country{USA}
}

\begin{abstract}
 Adaptive Virtual Reality (VR) systems have the potential to enhance training and learning experiences by dynamically responding to users’ cognitive states. This research investigates how eye tracking and heart rate variability (HRV) can be used to detect cognitive load and stress in VR environments, enabling real-time adaptation. The study follows a three-phase approach: (1) conducting a user study with the Stroop task to label cognitive load data and train machine learning models to detect high cognitive load, (2) fine-tuning these models with new users and integrating them into an adaptive VR system that dynamically adjusts training difficulty based on physiological signals, and (3) developing a privacy-aware approach to detect high cognitive load and compare this with the adaptive VR in Phase two. This research contributes to affective computing and adaptive VR using physiological sensing, with applications in education, training, and healthcare. Future work will explore scalability, real-time inference optimization, and ethical considerations in physiological adaptive VR.
  
\end{abstract}



\maketitle

\section{Introduction}
Virtual Reality (VR) is more than a technological advancement—it is a transformative medium that redefines how we perceive, interact with, and adapt to digital environments. VR provides an immersive, safe, and credible simulation of real-life training environments \cite{choi2015virtual, bailenson2018experience, bigonah2025systematic}. However, most current VR training applications provide unvaried training to all users and do not consider individual differences, such as cognitive abilities or emotional responses. This lack of adaptation can lead to cognitive overload, increased stress, and frustration, ultimately reducing learning efficiency and user engagement \cite{sweller2019cognitive, sandi2013stress}.

In my interdisciplinary PhD research, I explore the intersection of affective computing, VR design, and physiological sensing. Affective computing seeks to develop systems capable of recognizing, interpreting, and adapting to human emotions. At the same time, VR design provides the tools to craft experiences that can meaningfully engage and influence users. Physiological sensing bridges these fields, offering real-time insights into users' cognitive and emotional states through physiological signals such as heart rate variability (HRV) and eye-tracking \cite{souchet2022measuring, thayer2009heart, moon2023influence}.

High cognitive load has been shown to induce stress and frustration, which negatively impact user performance and motivation. When mental demands exceed cognitive resources, users experience a loss of control, leading to emotional distress \cite{paas2003cognitive}. Additionally, excessive cognitive demands activate the autonomic nervous system, increasing physiological arousal markers such as pupil dilation and reduced HRV \cite{laborde2017heart}. This can create a feedback loop where stress further impairs cognitive function, reducing engagement and learning outcomes \cite{sweller2019cognitive}.

To address this, adaptive VR environments dynamically adjust task difficulty and feedback mechanisms based on users' real-time cognitive and emotional states. By integrating Machine Learning (ML) models trained on physiological data, these systems can prevent cognitive overload, regulate stress, and enhance user experience \cite{shin2009evaluation, kritikos2021personalized}. Additionally, adaptive VR offers solutions for diverse applications, from education, which can adjust difficulty levels to optimize learning, to healthcare, and serve as a non-invasive, cost-effective alternative for pain management or stress reduction \cite{sweller2019cognitive}.

This research aims to develop a physiological adaptation framework based on cognitive load detection that adjusts VR training difficulty in real time. Specifically, this study:
\begin{itemize}
\item Develops a machine learning-based model using HRV and eye-tracking data to detect cognitive load and emotional stress.
\item Conducts a three-phase user study to train and fine-tune ML models using the Stroop task as ground truth.
\item Evaluates the efficiency of adaptive VR on user learning, engagement, and stress mitigation compared to privacy-aware adaptive and regular VR.
\end{itemize}

By addressing the interplay between cognitive load, emotions, and physiological responses, this research contributes to affective computing, adaptive VR, and human-centered design. 

\section{Related Work}
A systematic literature review was conducted using the Web of Science database, covering ACM, IEEE, and Springer publications from 2019 to 2024 to identify research gaps in affective computing, physiological sensing, and VR design. The search combined terms related to adaptation (e.g., "personalized," "adaptive") with biofeedback-related terms (e.g., "EEG," "GSR," "eye tracking") and "virtual reality." A total of 164 papers were identified, with 117 deemed relevant after the selection of the title and abstract. After filtering, 38 papers met the inclusion criteria, focusing on real-time physiological data use rather than post-analysis.

\subsection{Cognitive Load, Emotion, and Physiological Responses in VR}
Cognitive load significantly influences users' emotions, particularly stress and frustration. When cognitive demands exceed a user's working memory capacity, stress levels increase, reducing performance and engagement \cite{sweller2019cognitive, paas2003cognitive}. Physiological responses such as pupil dilation and reduced HRV correlate with high cognitive load, making them valuable signals for real-time adaptation \cite{laborde2017heart}. Additionally, cognitive overload can trigger emotional distress, leading to a feedback loop where increased stress further impairs cognitive function \cite{sweller2019cognitive}. Understanding these physiological responses allows computational models to detect cognitive load and dynamically adjust VR experiences to mitigate stress and enhance engagement.

\subsection{VR Adaptation Mechanisms}
VR adaptation mechanisms can be categorized into two primary approaches: \textbf{rule-based adaptation} and \textbf{data-driven adaptation}.

\textbf{Rule-Based Adaptation:} These approaches use predefined heuristics to modify VR parameters such as task difficulty, environmental features, and interaction mechanics. Some studies adjust difficulty dynamically to maintain engagement by modifying complexity \cite{blum2019heart, dey2019exploration, izountar2021towards}. Others adapt experience mechanics by altering users' virtual positions or modifying stimulus characteristics \cite{uyan2024cdms, apicella2024domain, baldini2024novel}. Environmental adaptations, such as adjusting colors or introducing elements like fog, also play a role in immersive adaptation \cite{gomes2021adaptive, gupta2024caevr, prabhu2020biofeedback}.

\textbf{Data-Driven Adaptation:} More recent approaches integrate physiological and behavioral data to adapt experiences. These models leverage physiological signals such as heart rate, EEG, and eye tracking to infer user state and modify the environment accordingly. For instance, Baldini et al. \cite{baldini2024novel} developed a VR system that mitigates cybersickness by adapting navigation speed and scene complexity based on EEG signals. Some studies use NPC behavior adaptation, such as streamlining NPC interactions based on user responses \cite{chiossi2024optimizing, chiossi2022virtual}. While these adaptation mechanisms enhance immersion, most rely on predefined heuristics. To enhance adaptability, computational models that leverage real-time physiological sensing have been explored.

\subsection{Computational Models for Adaptive VR}
Computational models for adaptation in VR fall into two primary categories: \textbf{direct physiological adaptation} and \textbf{ML-based adaptation}. 

\textbf{Direct Physiological Adaptation:} Some studies use real-time physiological signals to trigger immediate environmental changes. For example, changes in electrodermal activity (EDA) can adjust lighting or soundscapes in VR relaxation scenarios \cite{salminen2024meditating, moldoveanu2023immersive}. Other systems use gaze behavior to provide adaptive hints in training applications \cite{drey2020towards}.

\textbf{ML-Based Adaptation:} More advanced models use machine learning to predict user states and adapt experiences. However, few studies have implemented real-time physiological adaptation using ML \cite{uyan2024cdms, apicella2024domain, prabhu2020analyzing, ibanez2021using, gomes2023use}. Existing models primarily focus on arousal detection but lack multimodal emotion recognition. Techniques such as artificial neural networks (ANN), support vector machines (SVM), and random forests have been used to extract features from physiological signals and optimize virtual environments \cite{heyse2019personalized, kritikos2021personalized, de2022adaptive, izountar2021towards}. Despite these advancements, real-time physiological adaptation remains an underexplored area, particularly in privacy-aware adaptive systems.

\subsection{Privacy and Ethical Concerns in Physiological Adaptation}
While physiological adaptation offers promising ways to enhance VR experiences, ethical concerns remain largely unaddressed. The collection and processing of real-time physiological data raise privacy concerns, particularly regarding data security, user consent, and potential misuse \cite{williams2022immersive, chiossi2022virtual}. None of the studies propose frameworks for privacy-aware physiological adaptation, leaving a critical gap in the field.

\textbf{Research Gap:} There is a need for adaptive VR systems that not only integrate physiological data for real-time adaptation but also prioritize privacy considerations. This research aims to develop a privacy-aware physiological adaptation framework that detects cognitive load and dynamically adjusts VR training experiences in real time.

\section{Research Goals and Methodology}
The first goal is to develop machine learning models using eye-tracking and heart rate data to detect stress caused by high cognitive load. The VR system then adapts the experience to maintain an optimal cognitive state. Another goal is to create privacy-aware physiological sensing frameworks that respect user data while ensuring adaptability. The final step is to evaluate the effectiveness of adaptive VR in preventing high cognitive load and improving learning. This research seeks to answer the following questions, and the next sections are the approaches to solving them. 

\textbf{Research Questions:}
\begin{itemize}
\item RQ1: What physiological signals and computational models are currently used to detect and classify users’ emotional states in VR?
\item RQ2: What design principles should guide the development of user-centered physiological adaptive VR systems?
\item RQ3: How do privacy-aware cognitive load detection methods impact the effectiveness of adaptive VR systems compared to direct physiological signal-based approaches?
\end{itemize}

These research questions follow a three-phase approach:

\subsection{Phase 1: Cognitive Load Detection Using Physiological Signals}
\begin{figure*}[htb]
    \centering
    \includegraphics[width=0.5\textwidth]{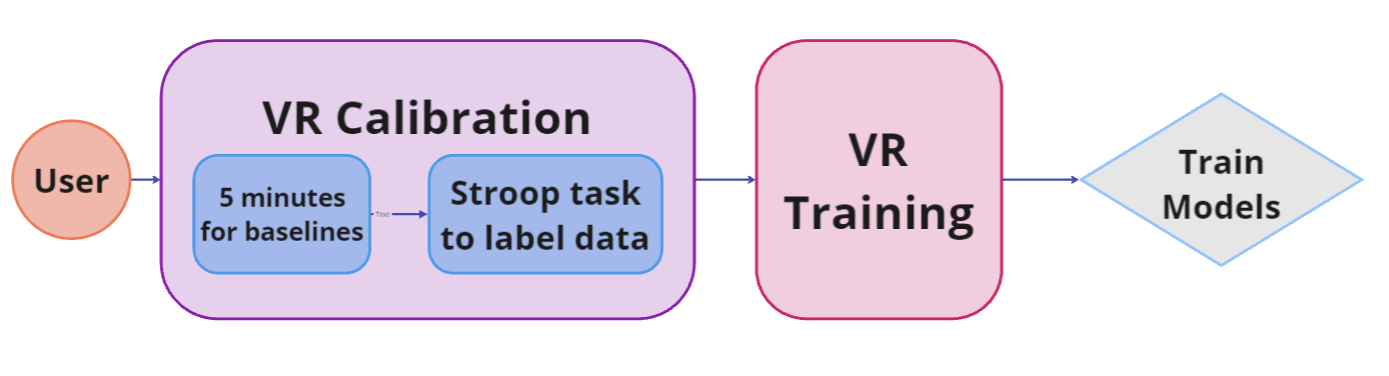}
    \caption{\small Phase 1: Initial data collection pipeline where users undergo VR calibration, including baseline measurements and the Stroop task, followed by VR training to collect physiological signals for machine learning model training.}
    \label{fig:phase1}
\end{figure*}

The first phase involves conducting a user study to collect physiological data and establish cognitive load as ground truth using the \textit{Stroop task}. First, in a calibration room, participants’ baselines will be recorded \cite{kreibig2010autonomic}. Then, participants’ physiological responses will be recorded while completing the Stroop task, and supervised learning models will be trained based on these labeled data. Next, participants enter the VR training, where we collect their eye tracking and heart rate. After the user study, we analyze the data and train our ML model to predict high and low cognitive load (Fig.~\ref{fig:phase1}.)
Stroop task is widely used in cognitive research due to its ability to induce controlled levels of mental workload \cite{stroop1935studies}. Alternative techniques, such as n-back tasks, also measure cognitive load but focus more on working memory rather than response inhibition and attentional control \cite{miller2009n}.

Physiological signals, including \textit{eye tracking and heart rate variability}, will be collected during the study. Eye tracking provides insights into cognitive processing through metrics like pupil dilation, fixation duration, and fixation count \cite{eckert2021cognitive}. Increased pupil dilation and prolonged fixations are linked to higher cognitive load \cite{lee2023measuring}. HRV is a well-established physiological marker of autonomic nervous system activity and cognitive effort \cite{thayer2009heart}. Decreased HRV correlates with stress and high workload, making it a reliable indicator for adaptation in VR environments \cite{laborde2017heart}.

\subsection{Phase 2: Machine Learning-Based Real-Time Adaptation}
\begin{figure*}[htb]
    \centering
    \includegraphics[width=0.5\textwidth]{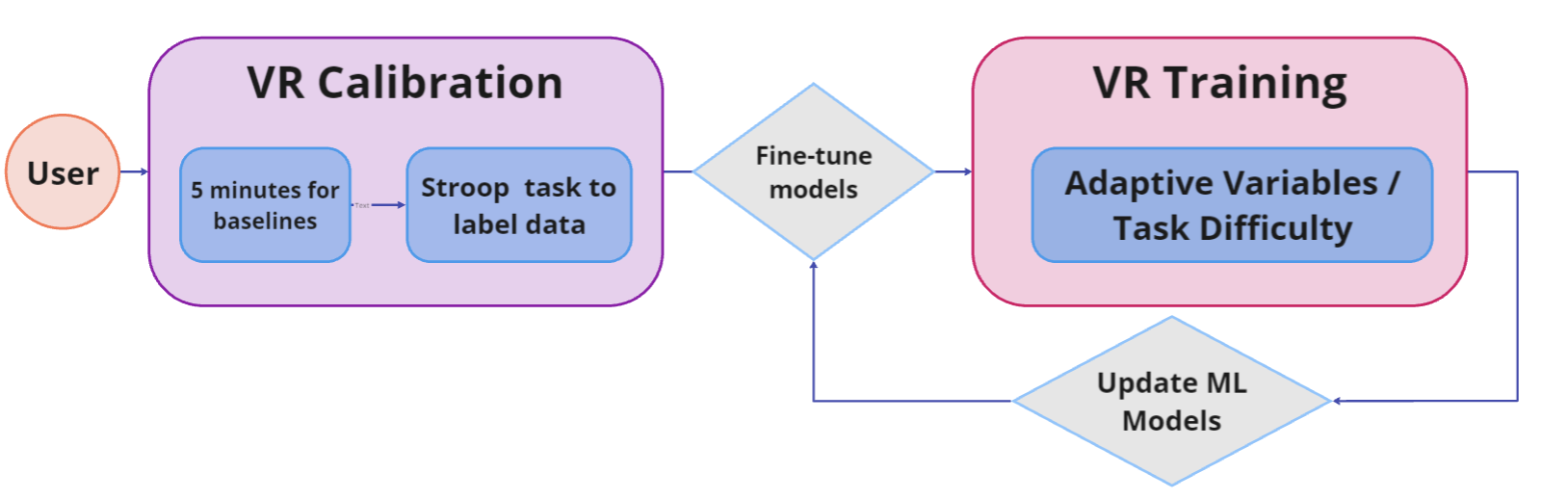}
    \caption{Phase 2: Adaptive VR system where trained models are fine-tuned, real-time task difficulty is adjusted based on cognitive load detection, and machine learning models are updated iteratively for improved adaptation.}
    \label{fig:phase2}
\end{figure*}

In the second phase, the trained models from Phase 1 will be fine-tuned using Stroop task data from new users. Once refined, these models will be integrated into the adaptive VR system, where real-time physiological signals will dynamically adjust task difficulty. Participants will be assigned to either an adaptive VR or regular VR group. The adaptive VR system will continuously adjust based on cognitive load detection, while the regular group will manually request hints. To evaluate the system, we will use other measures, such as NASA-TLX to assess overall mental workload, presence questionnaire to measure immersion, and a system usability scale for user experience evaluation. Performance metrics will include task completion time, error rate, and VR logs, such as the number of hints requested. Data collection will involve both physiological and self-reported measures to ensure a comprehensive evaluation.
Given the success of Multilayer Perceptron and Random Forest in prior work \cite{nasri2024exploring}, these models will be used for classification. RF is known for its robustness with small datasets and high interpretability \cite{breiman2001random}, while MLP captures complex non-linear relationships in physiological data \cite{lecun2015deep}. Performance will be evaluated using accuracy, F1-score, and AUC-ROC. The models will be fine-tuned with Stroop task data from new users before being deployed in the adaptive VR system where training difficulty dynamically adjusts based on cognitive load (Fig.~\ref{fig:phase2}.)

\subsection{Phase 3: Privacy-Aware Cognitive Load Detection}
The third phase of this research focuses on developing and evaluating a privacy-aware approach for cognitive load detection in adaptive VR systems. While eye tracking is a powerful tool for inferring cognitive states, the raw data—such as exact gaze coordinates, scanpaths, and fixation patterns—can inadvertently reveal sensitive personal information, including emotional responses, interests, intent, and even medical conditions. This raises concerns about user consent, data ownership, and the potential misuse of identifiable behavioral patterns.
To address these concerns, this phase will explore feature abstraction methods that reduce privacy risks while maintaining model performance. For instance, we will investigate the use of \textit{gaze entropy}, which provides a higher-level summary of gaze behavior without storing raw gaze coordinates, alongside other aggregate features such as fixation counts and dwell time distributions. These features aim to protect user privacy by making it harder to reverse-engineer identity or intent from the data.
In this phase, a user study will be conducted using the Stroop task to label cognitive load data, as in previous phases. Participants’ physiological responses will be recorded during task completion to establish ground truth. The collected data will then be used to compare the performance of cognitive load detection models that use privacy-aware features against those that rely on direct raw eye tracking data.
We will deploy both models in two versions of an adaptive VR system and evaluate their effectiveness based on cognitive load classification accuracy, user performance metrics (completion time, error rates, number of hints), and subjective workload assessment (NASA-TLX). The goal is to determine whether the privacy-aware model can maintain comparable accuracy in detecting cognitive load and adapting the VR experience, thereby balancing personalization and ethical data use.

\section{Results to Date}
To date, I have co-designed and evaluated VR environments for advanced manufacturing training, particularly in cold spray technology \cite{nasri2024designing}. In another effort, I developed ML models using eye tracking metrics, including pupil dilation and fixation duration, to classify mental workload \cite{nasri2024exploring}. The dataset consisted of 19 participants, with 10 labeled as experiencing high cognitive load and 9 as low, based on the mental demand subsection of the NASA-TLX. Although the class distribution was close to balanced, we used 3-fold cross-validation during model training to reduce the risk of overfitting and ensure a robust evaluation of model performance. Our results indicate that the MLP model achieved an accuracy and precision of 0.84, indicating that the RF model reached an accuracy of 0.72 and a precision of 0.73, suggesting reasonable performance but with some overfitting compared to the MLP model. More details are depicted in Table~\ref{table}.

\begin{table} [htb] 
    \centering
    \caption{Performance of MLP and RF models for predicting cognitive load using fixation duration and mean pupil dilation. The dataset included 19 participants (10 high, 9 low cognitive load). Although the class distribution was nearly balanced, we used cross-validation to reduce the risk of bias in performance estimates.}
    \resizebox{\linewidth}{!}{%
    \begin{tabular} { | l | c | c | c | c |} 
    \hline
     \textbf{Model} & \textbf{Accuracy} & \textbf{Precision} & \textbf{Recall} & \textbf{F1}\\
    \hline 
         MLP & 0.84 & 0.84 & 0.94 & 0.88 \\
     \hline
   RF & 0.72  & 0.73  & 0.90 & 0.81 \\
    
\hline
\label{table}
    \end{tabular}
    }
\end{table}

Currently, I am developing a privacy-aware approach leveraging eye tracking and gaze data to classify cognitive load. In this approach, I am implementing gaze entropy \cite{shiferaw2019review} alongside user behavior metrics (e.g., completion time and error rate) to classify users with high and low cognitive load.

\section{Next Steps and Long-Term Goals}
Currently, my dissertation is in the experimental phase, finalizing physiological sensing frameworks and ML models. The goal is to contribute to privacy-aware adaptive VR, affective computing, and real-time interaction adaptation. This research aims to provide a scalable adaptive framework that can be applied beyond academic research to industrial training, stress management, and clinical interventions. If validated, the models developed here could improve personalized learning environments, cognitive rehabilitation, and mental health applications in VR.

\subsection{Milestones with Success Criteria and Dissemination Plan}
\begin{itemize}
\item \textbf{Spring 2025:} Finalize VR environment development, conduct Phase 1 user study, collect physiological data (\textit{N=25 participants, ensuring at least 90\% usable physiological recordings, submit findings to CHI and ETRA}).
\item \textbf{Summer 2025:} Train ML models for cognitive load detection (\textit{targeting at least 80\% accuracy on unseen participants, submit findings to IEEE VR, UMAP}).
\item \textbf{Fall 2025:} Pass PhD proposal defense, refine adaptation strategies, and develop adaptive VR, conduct Phase 2 user study for adaptive VR effectiveness.
\item \textbf{Spring 2026:} Conduct Phase 3 user study for privacy-aware adaptive VR effectiveness(\textit{analyzing user performance gains and cognitive load between groups, submit results to UMAP, CHI}).
\item \textbf{Summer 2026:} Analyze results and fine-tune ML models (\textit{targeting acceptance at a top-tier venue such as CHI, UMAP}).
\item \textbf{Fall 2026:} Write and defend dissertation, develop recommendations for scaling adaptive VR in real-world applications.
\end{itemize}

This PhD consortium will provide invaluable feedback on my research direction, help me refine methodologies, and connect me with user modeling experts in VR and machine learning. Participating in ACM UMAP 2025 will enhance the impact of my work and ensure that my contributions align with cutting-edge research in physiological adaptive systems.

\subsection{Supervisors and PhD Program}
I am conducting this research under the supervision of Dr. Casper Harteveld and Dr. Leanne Chukoskie at Northeastern University, Boston, Massachusetts. I am a PhD student at the College of Art Media and Design in the Interdisciplinary Media and Design program, which I am interdisciplinary between Computer Science and VR Design. My research has been funded by NSF grant focusing on developing adaptive learning extended reality system and intelligent tutoring systems.

\bibliographystyle{ACM-Reference-Format}
\bibliography{ref}

\end{document}